\documentclass[sigconf]{acmart}

\usepackage{makecell}

\AtBeginDocument{%
  \providecommand\BibTeX{{%
    \normalfont B\kern-0.5em{\scshape i\kern-0.25em b}\kern-0.8em\TeX}}}

\setcopyright{acmlicensed}
\copyrightyear{2023}
\acmYear{2023}

\author{Ivan Flechais}
\affiliation{%
  \institution{University of Oxford}
  \country{United Kingdom}}

\email{ivan.flechais@cs.ox.ac.uk}
\orcid{0000-0002-3620-0843}

\author{George Chalhoub}
\affiliation{%
  \institution{University College London \\
  \& University of Oxford}
    \country{United Kingdom}}

\email{g.chalhoub@ucl.ac.uk}
\orcid{0000-0003-2082-2610}

\acmConference[NSPW '23]{New Security Paradigms Workshop}{September, 2023}{Segovia, Spain}
\begin{document}


\title{Practical Cybersecurity Ethics: Mapping CyBOK to Ethical Concerns}


\renewcommand{\shortauthors}{Chalhoub and Flechais}

\begin{abstract}
Research into the ethics of cybersecurity is an established and growing topic of investigation, however the translation of this research into practice is lacking: there exists a small number of professional codes of ethics or codes of practice in cybersecurity, e.g. the ISSA or the UK Cyber Security Council's code of ethics, however these are very broad and do not offer much insight into the ethical dilemmas that can be faced while performing specific cybersecurity activities. In order to address this gap, we leverage ongoing work on the Cyber Security Body of Knowledge (CyBOK) to help elicit and document the responsibilities and ethics of the profession. 

Based on a review of the existing literature on the ethics of cybersecurity, we use CyBOK to frame the exploration of ethical challenges in the cybersecurity profession through a series of 15 interviews with cybersecurity experts. Our approach is qualitative and exploratory, aiming to answer the research question ``What ethical challenges, insights, and solutions arise in different areas of cybersecurity?''. Our findings indicate that there are broad ethical challenges across the whole of cybersecurity, but also that different areas of cybersecurity can face specific ethical considerations for which more detailed guidance can help professionals in those areas. In particular, our findings indicate that security decision-making is expected of all security professionals, but that this requires them to balance a complex mix of different technical, objective and subjective points of view, and that resolving conflicts raises challenging ethical dilemmas. We highlight our participants' concerns about the growing use of AI technology in cybersecurity, and discuss the implications of applying AI to decision-making.  

We conclude that more work is needed to explore, map, and integrate ethical considerations into cybersecurity practice; the urgent need to conduct further research into the ethics of cybersecurity AI; and highlight the importance of this work for individuals and professional bodies who seek to develop and mature the cybersecurity profession in a responsible manner. 

\end{abstract}



\maketitle


\section{Introduction}
Cybersecurity professionals wield enormous power. In pursuing the goal of protecting computer systems, their responsibilities can variously lead them to monitor people and review sensitive information; investigate threat actors; document and prosecute insiders; subject others to attack through penetration tests or ``ethical hacking'' activities; decide on limiting, quarantining, revoking and denying access to systems and data in the face of ongoing attacks; or deal with the anguish, betrayal, and trauma arising from harmful cyber attacks. As noted by Christen et al. \cite{christen2020ethics}, \emph{``Overemphasising cybersecurity may violate fundamental values such as equality, fairness, freedom or privacy. However, neglecting cybersecurity could undermine citizens' trust and confidence in the digital infrastructure, in policy makers and in state authorities.''} Such power over other people's actions and freedoms should come with clear, transparent, and detailed ethical oversight, however this is far from the case in practice.

Research into the ethics of cybersecurity is an established and growing topic of investigation (see Sections 2.1 and 2.2), however the translation of this research into professional practice is lacking: there exists a small number of professional codes of ethics or codes of practice in cybersecurity, e.g. the ISSA or the UK Cyber Security Council's code of ethics, however these are very broad and do not offer much insight into the ethical dilemmas that can be faced while performing specific cybersecurity activities.

Significant efforts are underway to improve the maturity of cybersecurity: ranging from improvements in secure software development lifecycles, data protection law and regulation, cyber insurance, or efforts to codify the Cyber Security Body of Knowledge (CyBOK). In order to investigate the gap we identify between the research and practice of ethics in the cybersecurity profession, we propose the following research question \emph{``What ethical challenges, insights, and solutions arise in different areas of cybersecurity?''} Our approach is qualitative and exploratory: drawing on CyBOK and the principlist framework for cybersecurity ethics by Formosa et al. \cite{formosa2021principlist}, we engage with professionals in the field to elicit, map, and deepen our understanding of the key issues in specific areas of cybersecurity.

In Section \ref{background}, we review the current state of cybersecurity ethics in both research and professional practice. Section \ref{methodology} describes our research approach, and in Section \ref{findings} we present our results. We discuss these in Section \ref{discussion}, before concluding with suggestions for future work.

\section{Background}
\label{background}
The ethics of cybersecurity is establishing itself as a field of ethical research in its own right, sharing similarities with other fields such as bioethics, digital ethics, or the ethics of artificial intelligence, but having its own specific characteristics. While the ethics of cybersecurity has been dominated by concerns around privacy values, it is becoming clear that there are wider ethical challenges that arise \cite{10.1145/3555537,chalhoub_thesis_2022,10.1145/3544548.3581114,10.1145/3596267,10.5555/3488905.3488916,10.1007/978-3-030-50309-3_21,10.1145/3334480.3382850}. Christen et al. \cite{CHRISTEN20171} explore these and Figure 1 summarises some of the key relationships between cybersecurity values and other ethical values. 

In the following we start by presenting an overview of the underlying principles of cybersecurity ethics. We then explore different ethical frameworks for cybersecurity which aim to provide a more pragmatic structure to help evaluate different ethical questions. Then we provide a brief outline of the Cyber Security Body of Knowledge (CyBOK), and conclude by focusing on how professional codes of practice currently provide guidance for practitioners.

\begin{figure}
	\centering
	\includegraphics[width=8cm]{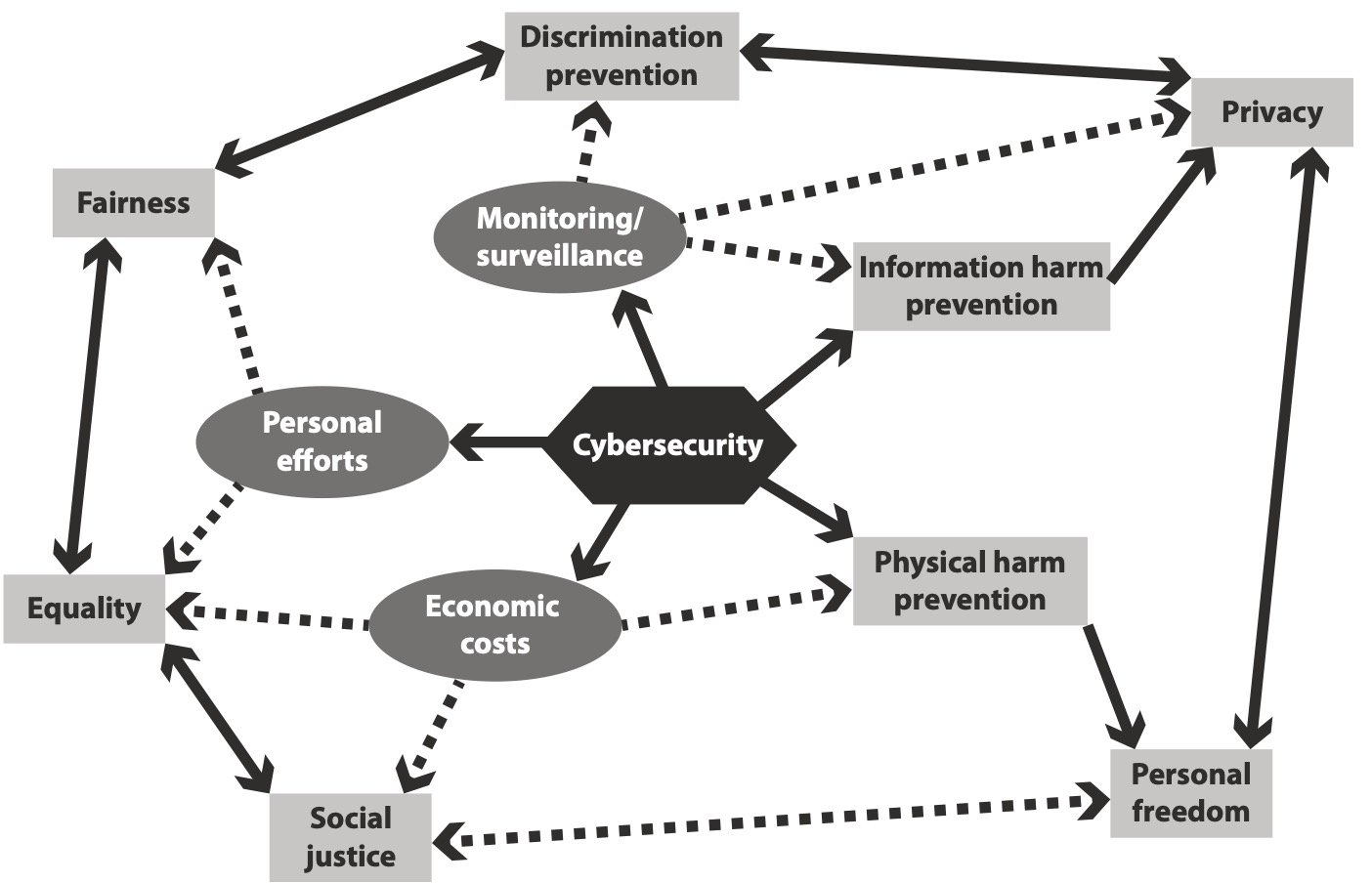}
	\caption{Value Conflicts in CyberSecurity. Arrows with continuous lines show positive (i.e., supporting) relations, whereas arrows with dotted lines show conflicting relations. Reproduced from \cite{CHRISTEN20171}}
	\label{valueconflicts}
\end{figure}

\subsection{CyberSecurity Ethical Principles} \label{ethicalprinciples}

There is a wide variety of different approaches looking at the ethics of cybersecurity \cite{291190}. Aiming to categorise these,  Macnish and van der Ham \cite{macnish2021ethical} argue that there are three broad approaches, calling the first ``bottom-up'' which examines detailed cases and identifies ethical issues arising from these. In contrast, the second is labelled ``top-down'' and focuses on ethical values or theories as a starting point which are then applied to the cybersecurity context. Macnish et. al. also identify a third approach which they call ``pragmatist'', which focuses on the practices of cyber security professionals. While they note that such approaches usually focus more on framing the values of information security (such as Confidentiality, Integrity, Availability) rather than identifying ethical issues, principles, or solutions, it is important to note that there are a variety of professional codes of practice that fall into this category (see Section \ref{codesofpractice}). 
Formosa et al. \cite{formosa2021principlist} argue for two broad categories instead. The first aims to apply established ethical theories (consequentialism, deontological ethics, and virtue ethics \cite{mouton2015necessity}) for which the ethical textbook by Manjikian \cite{manjikian2022cybersecurity} is a prominent example. The second category aims to outline a series of mid-level and domain-specific principles, an approach known as ``principlism''. In both categories, Formosa et al. \cite{formosa2021principlist} note that casuistry (an applied ethics approach that uses case-based reasoning to derive ethical insights) is widely used. 

Examining case studies of moral or ethical dilemmas in cybersecurity is a widespread approach that provides context and insight. Common case studies include generic example applications such as penetration testing or ethical hacking \cite{jamil2011ethical}, encryption in the context of privacy vs state surveillance (e.g. Rogaway \cite{rogaway2015moral} whose opening statement \emph{``Cryptography rearranges power''} neatly encapsulates a core issue), or electronic voting \cite{robinson2012ethical}. In contrast, other case studies focus on highly detailed and specific case studies, an example of which is the examination of the ENCORE project by Byers \cite{byers2015encore}. The ENCORE project  \cite{burnett2015encore} proposed a method to explore online censorship by harnessing cross-origin requests to covertly induce web browsers running on computers in various different countries into contacting specific websites and reporting back. Given (1) that this activity was non-consensual for the owners of those computers, (2) that there is significant potential harm to these owners arising from repressive regimes tracking attempts at accessing censored material, and (3) that the authors had already deployed and tested their approach in the wild, this research and its publication led to a significant ethical debate in the academic community. A key aspect of this debate centred around the fact that the research was approved by the authors' Institutional Review Board, who did not identify ethical issues arising from the technical details of the research. Following lengthy deliberation, the Program Committee for SIGCOMM voted to publish the paper, however they took the unprecedented decision to document their ethical concerns in a statement at the top of the paper \cite{byers2015encore}.

Whether drawing from case studies or applying ethical theories to the cybersecurity context, a key aim has been to identify and specify the core ethical values of cybersecurity. The chapter by van de Poel \cite{van2020core} provides a helpful overview of some of these, noting that ethical cybersecurity values can be clustered into aspects of security, privacy, fairness, and accountability. Christen et al. note in their introductory chapter that cybersecurity involves a balance between fundamental values such as equality, fairness, freedom or privacy and the need for protecting citizens' trust and confidence in the digital infrastructure, in policy makers, and in state authorities \cite{christen2020ethics}. 

\subsection{CyberSecurity Ethical Frameworks}
\label{ethicalframeworks}
Ethical frameworks aim to provide structure to help evaluate ethical questions. Loi and Christen \cite{loi2020ethical} provide a helpful overview of these and note that there are several notable ethical frameworks in cybersecurity.

\subsubsection{Human Rights Frameworks}
The first type of framework is based on human rights and how these are embedded in various legal and regulatory frameworks. Hildebrandt \cite{hildebrandt2013balance} discusses how rights can interact with cybersecurity, focussing specifically on privacy, data protection, non-discrimination, due process and free speech. Under EU law, these rights are protected, however cybersecurity activities (such as monitoring, profiling, filtering content) can come in conflict with these rights. Hildebrandt \cite{hildebrandt2013balance} argues that in the case where a conflict is resolved through a trade-off, infringing measures have to be balanced by effective safeguards. To achieve this balance, Hildebrandt draws on the triple test, derived from the second paragraph of Art. 8 of the European Convention of Human Rights, which requires that a right's infringement ``\textit{must be in accordance with the law, necessary in a democratic society and have a legitimate aim.}'' While a detailed review of various legal frameworks is beyond the scope of this paper, it is notable that ethical concerns have played a significant role in the formulation of EU data protection regulation, such as the General Data Protection Regulation, which enshrines seven principles (Lawfulness, fairness and transparency; Purpose limitation; Data minimisation; Accuracy; Storage limitation; Integrity and confidentiality (security); Accountability) and eight individual rights (The right to be informed; The right of access; The right to rectification; The right to erasure; The right to restrict processing; The right to data portability; The right to object; Rights in relation to automated decision-making and profiling).

\subsubsection{Theory of Contextual Integrity}
The second approach is Nissembaum's theory of Contextual Integrity \cite{nissenbaum2004privacy}. This theory has enjoyed widespread success arising from its characterisation of privacy violations as violations of social norms from the transmission of information between persons. Social norms are grounded in the specific contexts of each situation, allowing the exploration of privacy in a manner that is sensitive to societal, cultural, and wider contextual factors. 

\begin{figure*}
	\centering
	\includegraphics[width=14cm]{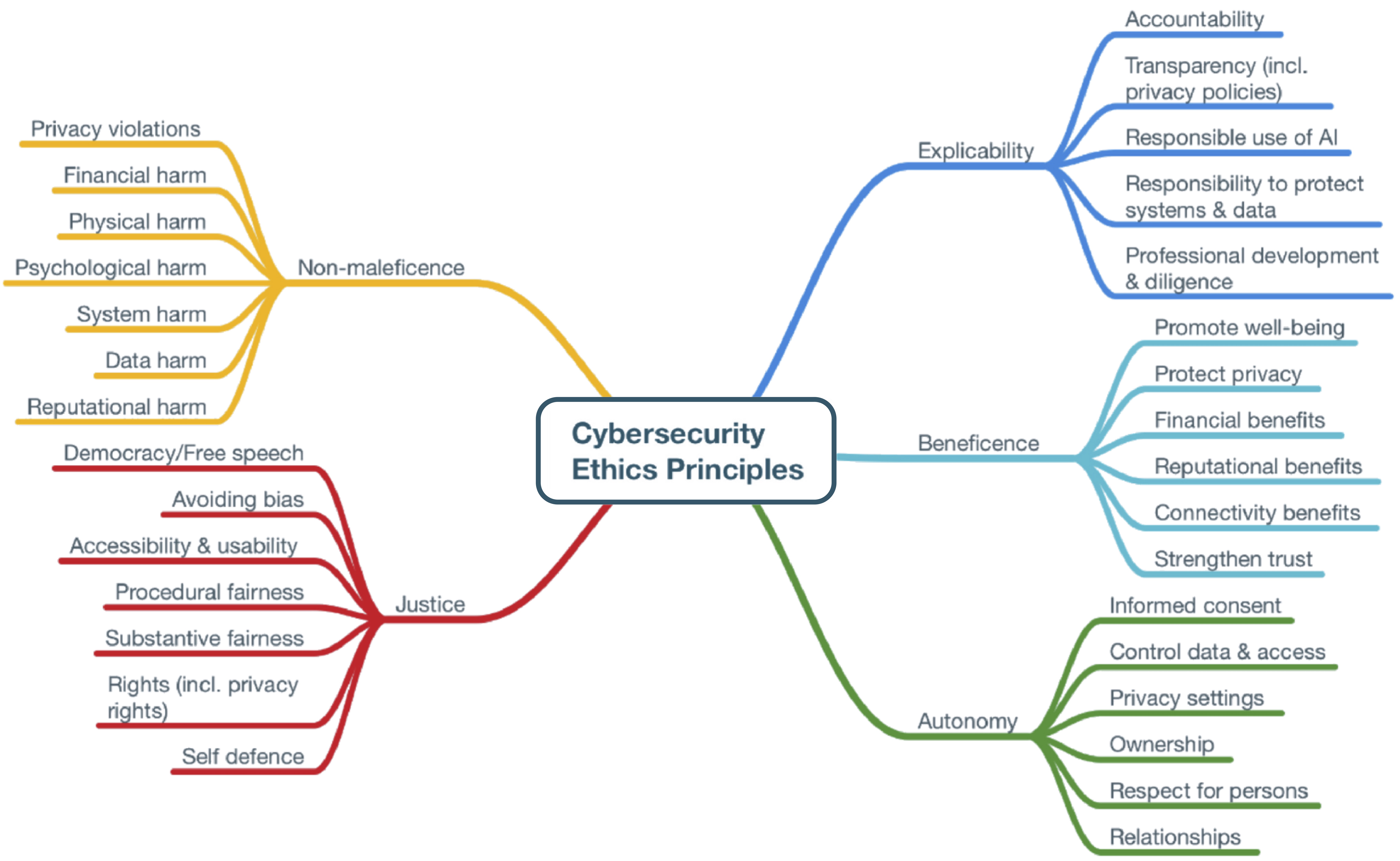}
	\caption{A principlist framework for cybersecurity ethics. Reproduced from \cite{formosa2021principlist} }
	\label{framework}
\end{figure*}

\subsubsection{Ethics of Risk}
The third type of framework explores the ethics of risk. Drawing on the significant body of work from Sven Ove Hansson on the ethics of risk \cite{hansson2013ethics}, Macnish and van der Ham \cite{macnish2021ethical} argue that cybersecurity is the inverse of risk. Where risk is defined as the likelihood of harm arising from a threat, security grows as risk reduces; conversely, as security decreases, the likelihood and impact of harm increases. In positioning cybersecurity in terms of risk, a number of interesting observations can be made. The first observation is that security is always forward-looking: since risk aims to anticipate the likelihood and impact of future threats, then likewise security is only concerned about the future -- should an incident occur, it is no longer in the realm of risk or security but in the realm of harm, crisis and security failures. The second observation was made by Herington \cite{herington2018contribution} who noted that security has subjective, objective, and affective dimensions: a person could be secure (objective), believe that they are secure (subjective), but not feel secure (affective). These qualities are interrelated in that they can influence one another, however they are distinct and security needs to satisfy a number of potentially conflicting perspectives: rational (objective), personal (subjective), and emotional (affective). 

Sven Ove Hansson and \cite{hansson2013ethics}, Macnish and van der Ham \cite{macnish2021ethical} also highlight that there are four notable ethical implications from looking at risk: (1) the distinction between objective and subjective harms, (2) the challenges of calculating probabilities, (3) the recognition of fallacies, and (4) the problems arising from risk thresholds and distribution. Evaluating harm (1) highlights the inherent quality of security having subjective, objective and affective facets, however most approaches tend to aim for objective measurements and eschew subjective or emotional dimensions. Calculating the probability of a future attack (2) also shares these sensitivities: in cybersecurity attackers are intelligent adversaries who may not follow patterns of previously seen behaviour. Consequently, calculating probabilities requires an element of subjective judgement, which raises questions when different opinions vary or conflict with objective data. There are a number of fallacies (3) that permeate cybersecurity \cite{hansson2004fallacies}, including the ``sheer size fallacy'' (\emph{if one risk is smaller than an acceptable unrelated risk, then it should be accepted}), the ``technocratic fallacy'' (\emph{since cybersecurity risks can be highly technical, only technical people can decide on what to do about them}), or the ``fallacy of pricing'' (\emph{since we have to weigh the costs of risks against their benefits, it is necessary to place a monetary value on risks}). Finally, it is important to note that risks are not tolerated equally, and neither are they fairly and evenly distributed (4): those who make decisions about risks may not be impacted by them or pay for the costs associated with the risk; a company's underinvestment in security may result in significant harm to others (e.g. their customers) and not directly to themselves; one business's estimate of a tolerable level of risk may be deemed unacceptable by a regulator; a poorly implemented authentication solution may cause usability and productivity impacts to users, etc.

\subsubsection{Principlist Frameworks}
The last consist of principlist frameworks, an example of which was used in the highly influential Menlo report \cite{bailey2012menlo}.  Principlism is a form of deontology, and principlist frameworks are articulated around a small number of fundamental principles derived from moral and professional ethical practices. These principles then drive what duties need to be satisfied, however complications arise when different duties are in conflict. A principlist framework thus aims to help navigate these issues by providing a lightweight means of helping to identify possible conflicts, however the exact nature, context, and importance of specific factors in these conflicts is left to the deliberation of researchers and practitioners. Furthermore, the resolution of such conflicts is also left open to the interpretation of the users of the framework. 
The Menlo report proposes four principles: Respect for Persons, Beneficence,
Justice, Respect for Law and Public Interest. The first three of these are drawn from the Belmont report \cite{united1978belmont} that focuses on the protection of human research subjects, and the fourth is proposed as an additional category to highlight the wider legal and public interest in cybersecurity.

Formosa et al. \cite{formosa2021principlist} noted the growing importance of Artificial Intelligence and associated ethical issues \cite{floridi2018ai4people} in the practice of cybersecurity. Drawing on the Menlo report and other principlist approaches \cite{van2020core,bailey2012menlo,loi2020ethical,weber2020cybersecurity,morgan2020care}, they propose a principlist framework to address ethical issues according to five different principles, the definitions of which are reproduced here verbatim: 
\emph{\begin{itemize}
	\item Beneficence: Cybersecurity technologies should be used to benefit humans, promote human well-being, and make our lives better overall.
	\item Non-maleficence: Cybersecurity technologies should not be used to intentionally harm humans or to make our lives worse overall.
	\item Autonomy: Cybersecurity technologies should be used in ways that respect human autonomy. Humans should be able to make informed decisions for themselves about how that technology is used in their lives.
	\item Justice: Cybersecurity technologies should be used to promote fairness, equality, and impartiality. It should not be used to unfairly discriminate, undermine solidarity, or prevent equal access.
	\item Explicability: Cybersecurity technologies should be used in ways that are intelligible, transparent, and comprehensible, and it should also be clear who is accountable and responsible for its use.
\end{itemize}}
This framework outlines different ethical values in cybersecurity, which can further be refined into more detailed concepts as illustrated in figure \ref{framework}. Given that this is the first framework to explicitly include the consideration of the ethics of AI in cybersecurity, we chose this approach to help frame the ethical aspect of our investigation as described in section \ref{methodology}.

\subsection{Cyber Security Body of Knowledge}
\label{cybok}
Cybersecurity is a concept that has been defined and characterised in a variety of different ways, most of which frame it as a process for protecting information by preventing, detecting, and responding to attacks \cite{nist_def}. Several properties of cybersecurity are regularly included in these definitions, the core three being confidentiality, integrity, and availability. Additional properties are sometimes added, such as authentication, non-repudiation, or utility, and the scope can also be wider than the protection of information to include computers, electronic communication systems, and electronic communications.

Going into more detail about the different areas of foundational and generally recognised knowledge that make up cybersecurity, the Cyber Security Body of Knowledge (CyBOK \cite{rashid2018scoping}) serves as a guide and maps the core elements of the discipline. At the time of writing, CyBOK has released v1.1 of its knowledgebase \cite{cybok}, which breaks down cybersecurity into five main categories: Human, Organisational \& Regulatory Aspects; Attacks \& Defences; Systems Security; Software and Platform Security; and Infrastructure Security. Within each of these categories, a total of 21 Knowledge Areas (KA) introduce and outline common material. Taken together, CyBOK can be used to understand the means and objectives of cybersecurity, mitigate against failures and incidents, and manage risks. Given that CyBOK provides a comprehensive breakdown of different areas of cybersecurity, we chose to use it to help frame our investigation as described in section \ref{methodology}. 

\subsection{Professional Codes of Practice} \label{codesofpractice}

A number of professional bodies have adopted codes of practice to help their members navigate the ethical challenges that can arise in the performance of their duties. The Association for Computing Machinery (ACM) with approximately 100,000 members (educators, researchers, and professionals) is the world's largest computer society, and has published a detailed Code of Ethics and Professional Conduct (\cite{acm}). This outlines and discusses the following general ethical principles:
\begin{enumerate}
    \item Contribute to society and to human well-being, acknowledging that all people are stakeholders in computing.
    \item Avoid harm.
    \item Be honest and trustworthy.
    \item Be fair and take action not to discriminate.
    \item Respect the work required to produce new ideas, inventions, creative works, and computing artefacts.
    \item Respect privacy.
    \item Honour confidentiality.
\end{enumerate}

More specifically to cybersecurity, the Information Systems Security Association (ISSA) has also published a code of ethics \cite{issa}, however this is quite short, consisting of the following:
\begin{itemize}
    \item Perform all professional activities and duties in accordance with all applicable laws and the highest ethical principles;
    \item Promote generally accepted information security current best practices and standards;
    \item Maintain appropriate confidentiality of proprietary or otherwise sensitive information encountered in the course of professional activities;
    \item Discharge professional responsibilities with diligence and honesty;
    \item Refrain from any activities which might constitute a conflict of interest or otherwise damage the reputation of or is detrimental to employers, the information security profession, or the Association; and
    \item Not intentionally injure or impugn the professional reputation or practice of colleagues, clients, or employers.
\end{itemize}

The UK Cyber Security Council has its own code of ethics for cybersecurity grounded on the principles of integrity, professionalism, and credibility. In addition, the council has made available some guidance for individuals, articulated around Fair Competition, Honesty, Inclusion, Integrity, Lawful behaviour, Professionalism, Reporting, and Competence. Complementing this, the council published 16 case studies which illustrate the core ethical considerations. In the same way that we use CyBOK to frame our research into the ethical challenges of cybersecurity, the scenarios presented by the UK Cyber Security Council are also categorised according to the five main Knowledge Areas of CyBOK.

The Forum of Incident Response and Security Teams (FIRST) has also developed its own guidance in the form of a code of practice for professionals involved in incident response \cite{ethicsfirst}. The code of practice outlines a number of duties (Duty of trustworthiness; Duty of coordinated vulnerability disclosure; Duty of confidentiality; Duty to acknowledge; Duty of authorization; Duty to inform; Duty to respect human rights; Duty to Team health; Duty to Team ability; Duty for responsible collection; Duty to recognize jurisdictional boundaries; Duty of evidence-based reasoning). In addition to providing a clear definition of these different duties, the code of practice also includes an appendix for dealing with dilemmas. This outlines how members may find themselves in a position where no action seems to satisfy all
of the ethical principles. To address such dilemmas, practitioners are encouraged to reflect on how stakeholders may be affected by their actions and to favour solutions that minimize the infringement of the Code.

Overall, the guidance and codes of practice offered by professional bodies are helpful, but they typically fall short of the detail and nuance that comes from the ethical work outlined in Sections \ref{ethicalprinciples} \& \ref{ethicalframeworks}. Moreover, we note that beyond mapping exemplar ethical case studies to different areas of cybersecurity, it is not clear what kinds of ethical questions can arise in the exercise of specific professional duties. As a result, our research aims to investigate this by exploring the kinds of ethical challenges, insights, and solutions that arise in different areas of cybersecurity. In the following section, we describe the methodology we used to investigate this.

\section{Methodology}\label{methodology}
Based on our review of the existing literature on the ethics of cybersecurity, we devised an interview guide to answer the research question: \emph{``What ethical challenges, insights, and solutions arise in different areas of cybersecurity?'' } 
We recruited 15 interviewees from outside our institutions through a mix of direct contact with existing connections and previous research participants who had consented to being contacted about future research. Our selection criteria required participants to have experience and knowledge of the cybersecurity profession.

We analysed the data using thematic analysis, which is an inductive coding process to help identify patterns in meaning from the data. Our analysis identified a number of themes and sub-themes pertaining to ethical concerns, experiences, and solutions that our participants related to different areas of cybersecurity.  

The study was ethically reviewed and approved by the Departmental Research Ethics Committee at our institution. 

\subsection{Recruitment}
We used several means to recruit our participants, including advertising on Twitter, Reddit, Mailing Lists and Blogs. We also reached out to participants on Slack channels and LinkedIn. To diversify our sample, we aimed to interview senior managers and executives who have likely made important security decisions. Since these are a hard-to-reach group \cite{faugier_sampling_1997}, we used the snowball sampling method \cite{goodman_snowball_1961} to recruit some participants, and worked with a consultant advisor who had wider access to senior executives working in security. We note that the results of the convenience sampling cannot be generalised to the target population because of the potential bias of the sampling technique \cite{BORNSTEIN2013357} and cannot be used to identify differences of population subgroups \cite{BORNSTEIN2013357}.

All of our participants were working at different companies. At the time of recruitment, interested participants were employees who were active at their company.

We asked interested participants to complete an online screening questionnaire.  We received 85 complete responses. In addition to asking demographic questions, we asked participants to provide details on their employment as well as their company size. 

We selected participants based on those who could best inform our research question and enhance our understanding of the ethical challenges, insights, and solutions arising in cyber security. Hence, we chose participants based on occupation (e.g. technical, managerial), field (e.g. malware, privacy, web security), relevance to CyBOK's knowledge base (e.g. human, organisational, and regulatory aspects), experience level (e.g. junior, senior), and diversity (e.g. gender, ethnicity). We note that all our participants worked in the private sector and none worked directly for the government or government affiliated employers.

We describe the demographics of our participants in Table \ref{table:demographics}.


\subsection{Pilot Study}
To validate our initial interview questions, we conducted a pilot study with two individuals in our research institution. We recruited the pilot participants through snowball sampling. Two researchers analysed the pilot interviews. We used the findings to identify potential problems (e.g., adverse events, time) in advance prior to conducting the full-scale study. We  didn't use the results from the pilot interviews, but we have refined our interview questions to ensure they are non-leading and clearer for participants. 

\subsection{Demographics}
Table \ref{table:demographics} summarises the demographics of our sample (n=15). We interviewed nine male and five female participants. Ages ranged from 18 to 55. Eight participants were interviewed in person, and seven remotely.

Additionally, we collected more contextual information about our participants and the business sector in which they operate (socioeconomic status of customers, types of products offered, geographic locations served). We conducted 5 interviews in-person (in secure locations in our institution) and 10 interviews remotely (on Microsoft Teams).

\begin{table}[h]

\centering
\footnotesize

\caption{Study Demographics}
 \label{table:demographics} 

\begin{tabular}{|c|l|c|c|c|c}
\hline
P\# & Age (M/F) & Occupation 
 & Cyber Security Field & CyBOK \\ \hline
P01 & 50-55 (M) & Managing Director &   \makecell{Privacy \& \\ Online Rights} &  \cite{cybok-por}  \\ \hline
P02 & 45-50 (M) & Systems Security Lead & \makecell{Secure Software\\ Lifecycle}  &  \cite{cybok-ssl}   \\ \hline
P03 & 45-50 (F) & Security Manager &  \makecell{Risk Management  \\ \& Governance } &  \cite{cybok-rmg} \\ \hline
P04 & 35-40 (M) & Security Engineer &   \makecell{Security Operations \& \\Incident Management}   &  \cite{cybok-soim}  \\ \hline
P05 & 40-45 (M) & Cloud Security Lead & \makecell{Network \\ Security } &  \cite{cybok-ns} \\ \hline
P06 & 40-45 (F) & Infosec Lead & \makecell{Web \& Mobile \\ Security}  &  \cite{cybok-wam} \\ \hline
P07 & 30-35 (M) & Security Engineer & \makecell{Malware \& Attack \\  Technologies} &  \cite{cybok-mat}  \\   \hline
P08 & 30-35 (F) & Freelance Ethical Hacker  & \makecell{Adversarial \\ Behaviours}  &  \cite{cybok-ab} \\ \hline
P09 & 45-50 (M) & Product Security Lead & \makecell{Forensics \& \\ Software Security} &  \cite{cybok-f,cybok-ss} \\ \hline
P10 & 18-25 (M) & Security Developer &  \makecell{Malware \& \\ Attack Technologies}  &  \cite{cybok-mat} \\ \hline
P11 & 30-35 (M) & Security Engineer & \makecell{OS  \& Virtualisation \\ Security} &  \cite{cybok-osv} \\ \hline
P12 & 25-30 (F) & Penetration Tester & \makecell{Web \& Mobile \\ Security} &  \cite{cybok-wam}  \\ \hline
P13 & 35-40 (F) & Security Manager & \makecell{Authentication  \\ \& Authorisation} &  \cite{cybok-aaa} \\ \hline
P14 & 30-35 (M) & Security Consultant & \makecell{Physical Layer \& \\ Telecom Security} &  \cite{cybok-plt} \\ \hline
P15 & 25-30 (M) & Privacy Engineer & \makecell{Human Factors \& \\ Law \& Regulation} &  \cite{cybok-hf,cybok-lr} \\ \hline

\end{tabular}

\end{table}

\subsection{Procedure}
\subsubsection{Semi-structured Interviews}
We followed a semi-structured interview protocol utilising an interview guide to maintain direction while keeping the interview open for both depth and breadth of topic exploration. 
In order to help focus our questions on specific aspects of cybersecurity professions, we made use of CyBOK's five broad categories:
\begin{enumerate}
	\item Human, Organisational \& Regulatory Aspects (e.g. risk management \& governance, law \& regulation, human factors, privacy \& online rights)
	\item Attacks \& Defences (e.g. malware \& attack technologies, adversarial behaviours, security operations \& incident management, forensics)
	\item Systems Security (e.g. cryptography, operating systems \& virtualisation security, distributed systems security, formal methods for security, authentication, authorization \& accountability)
	\item Software and Platform Security (e.g. software security, web \& mobile security, secure software lifecycle)
	\item Infrastructure Security (e.g. applied cryptography, network security, hardware security, cyber physical systems, physical layer \& telecommunications security) 
\end{enumerate}
In addition, to help focus on different aspects of cybersecurity ethics, we used the five principles proposed by \cite{formosa2021principlist}, outlined above in Section \ref{ethicalframeworks}.
Our interview guide is included in Appendix A.

\subsubsection{Thematic Analysis} \label{thematicanalysis}
The interview data was analysed using Thematic Analysis. According to \cite{cooper2012apa}, it is a common method of analysis in qualitative research and involves identifying, analysing, interpreting, and reporting patterns of meaning (known as themes or codes) from qualitative data. Thematic analysis is also frequently used with existing theoretical frameworks to provide interpretive power. Given that our approach made use of the principlist framework for cybersecurity ethics proposed by \cite{formosa2021principlist}, thematic analysis proved highly suitable in this regard, and helped to provide greater insight into detailed ethical considerations.

Two researchers were involved in the data collection and analysis. The primary researcher, who conducted most of the interviews, did an initial coding of the interview transcripts. To ensure credibility of the codes, a second researcher cross-checked all the codes against the interview transcripts. Any differences and/or issues arising from the coding were discussed and resolved among the two researchers. A codebook consisting of 122 codes emerged from the initial coding. These codes were then applied across other interviews through constant comparison, while new codes were added as they emerged and were deemed necessary. In further analysis, the researchers discussed and grouped the codes into themes. Regular coding meetings were held to discuss any emerging codes and to group the codes into families.

We observed data saturation~\cite{seale_quality_1999, corbin_basics_2014, guest_how_2006} between the 13th and the 15th interview; i.e. no additional issues or insights emerged from data and all relevant conceptual categories had been identified, explored, and exhausted. Hence, we stopped interviewing.

\subsection{Research Ethics} 
The University of Oxford's Central University Research Ethics Committee (CUREC) reviewed and approved the study (C1B-23HT-COML-003). Prior to each interview, participants were briefed and signed an informed consent form explaining our study and data confidentiality practices. Due to the sensitivity of our interviews, we asked participants not to name specific people or sites so that the interviews will be anonymous to some degree.

All interviews were AES 256 encrypted and stored in a physical safe in our organisation. Participants were thanked for their time with GBP \pounds 50 in electronic store vouchers. In addition, participants were reimbursed for out-of-pocket expenses related to participation, including travel, meals, accommodation, and childcare. Participants could withdraw themselves and their data at any point, without loss of compensation, and without providing a reason. No participant withdrew.

\section{Findings}
\label{findings}

\noindent In this section, we detail the findings of our study. We discuss our key findings organised according to the main themes of our analysis. The main themes are:
\begin{itemize}
\item Human, Organisational, Regulatory Aspects (\S\ref{R1})
\item Attacks \& Defences (\S\ref{R2})
\item Systems Security (\S\ref{R3})
\item Software \& Platform Security (\S\ref{R4})
\item Infrastructure Security (\S\ref{R5})
\end{itemize}

\subsection{Human, Organisational \& Regulatory Aspects}\label{R1}
Our participants reported a number of ethical concerns, dilemmas and challenges in relation to the Human, Organisational and Regulatory Aspects of cyber security which encompasses Risk Management and Governance, Law and Regulation, Human Factors and Privacy and Online Rights. Participants experience ethical challenges related to the constant need for maintaining confidentiality (\S \ref{R1.1}), balancing the competing interests of protecting their company's reputation and maintaining user security (\S \ref{R1.2}), and disclosing security risks without making users feel insecure (\S \ref{R1.3}). 

\subsubsection{Always Maintaining Confidentiality} \label{R1.1}
Our participants (n=5) stated that due to the nature of their profession, they handle and see private, sensitive, and  proprietary information that must be kept entirely secret -- adding that maintaining confidentiality is highly critical. They are instructed to maintain the confidentiality of information they come upon, or face significant ramifications for their career. However, they experienced ethical dilemmas and challenges deciding between maintaining or breaking confidentiality. Participants struggled to carefully weigh the ethical implications and informed decisions of whether or not to disclose confidential information. 
 Our participants stated this is a complex challenge that is dependent on many factors such as whether there is knowledge about harm, crime or illegal activities, and whether the confidential information should be disclosed to law enforcement or other government agencies. For instance, participant P07, who works at a cybersecurity and anti-virus company gave more insights on the ethical challenges to maintaining confidentiality while repairing their client's machines: 

\begin{quote}``\textit{We are asked to fix people's computers all the times. Our job is getting rid of the virus, not looking at our clients' photo albums or tax returns. We don't care what data is on their computer, we're only here to repair it. I'm not saying we do, but if we ever see any confidential information, I assume it can get quite complicated. If someone is cheating on their spouse or avoiding taxes, then it's none of our business. But if we find by accident that there is threat to human life or a child in danger, we have an obligation to report it.}''-- P07, Security Engineer \end{quote}

\subsubsection{Conflicts between Business \& Security Practices}  \label{R1.2}
Our participants (n=7) reported that conflicts of interest frequently arise in cyber security firms due to the competing interests between business and cyber security practices. 
They reported that conflicts of interest emerge between individual interests, public interests and corporate interests. Some of our participants suggested that cyber security practices should be prioritised over business or profit-making activities, as they strengthen the company's reputation and lead to trust over the long term. A common topic of discussion was the ethical challenges of dealing with data breaches: some companies tend to be reluctant to inform the public of data breaches in an attempt to protect their brand.  For instance, Participant P01, who manages a company that specialises in data privacy management tools, provided more insights on the conflicts of interests arising between data breaches and reputational damage: 

\begin{quote} \textit{``We had companies where they've had multiple breaches, seriously embarrassing breaches. The company is still trading and still has actually gone from strength to strength. When they had the breach, all they did was lower their prices for a while and then bring it up again. They actually increased their market reach. CEOs will then tell you: ``We got a plan to deal with it if it occurs''. They don't really care about the breach. They obviously care about the reputational damage. They've got a plan on how to deal with the reputational damage. They're actually not addressing the issue.''} -- P01, Managing Director \end{quote}

Moreover, Participant P03 who is manager at a small cyber security firm stated that a key ethical challenge is whether to conduct business with clients who are not willing to invest in adequate cyber security solutions.

\subsubsection{ Disclosing Security Risks Without Making Users Feel Insecure}  \label{R1.3}

Our participants (n=4) reported ethical challenges, and dilemmas in disclosing security risks without making their customers or users feel insecure. Our participants reported that this can be a challenging balancing act. On one hand, disclosing security risks is a transparent business practice that improves transparency and ensures that users take informed decisions. On the other hand, disclosing security vulnerabilities can risk making users feel unnecessarily insecure or anxious about their security. For instance, Privacy Engineer P15 provided more details on this challenge: \begin{quote} \textit{``Sometimes you don't want to put the end-user in a state where they're not so sure about the systems and how their data is being processed and stuff just because there's a minor issue which you are able to pick on your own. So it's very hard sometimes when you're looking at the business side as well as the ethical side.''} -- P15, Privacy Engineer \end{quote}

\subsection{Attacks \& Defences}\label{R2}
Our participants reported a number of ethical concerns, dilemmas and challenges in relation to the Attack \& Defence aspects of cyber security which encompasses Malware \& Attack Technologies, Adversarial Behaviours, Security Operations \& Incident Management and Forensics. Participants experience ethical challenges related to ethical security hacking and cyber intrusion (\S \ref{R2.1}), and defending against cyber attacks (\S \ref{R2.2}).

\subsubsection{Ethical Security Hacking and Cyber Intrusion} \label{R2.1}
Our participants (n=3) reported concerns and dilemmas in relation to ethical security hacking and cyber intrusion. Some of our participants were ethical hackers hired by different companies to test their security systems and find security vulnerabilities. They were responsible for applying offensive cyber tools and techniques to identify and drive security improvements. In return, they would reveal vulnerabilities to their client and create guidelines on how to address them, thus helping to secure company networks to protect trade secrets and business practices. Our participants reported that their most ethical concerns or challenges result from the possibility of finding illegal activity or unethical practices connected to their client. For instance, participant P08, a freelance ethical hacker who delivers penetration testing and red teaming capabilities for companies, discussed the ethical ramifications of the possibility of discovering illegal activity during a penetration test. 

\begin{quote} ``\textit{During a pentest, you could come across illegal activity. This is where it gets messy. My clients tend to be from all over the world, what is legal in the UK might be illegal in other countries. Do you report it to the authorities? You could get investigated yourself. If you report it, are you breaking your NDA? That all depends on the jurisdiction. Do you not report it? You could be complicit to a crime you don't report. I am personally aware of pentesting companies that have a provision allowing them to report illegal activities observed.}'' -- P08, Freelance Ethical Hacker \end{quote} 

Moreover, participant P12, who is a penetration tester, stated that they evaluate their own values and moral principles; and don't take any projects that could create a conflict with their own or with other societal ethical values.

While this finding is similar to the one reported in Section \ref{R1.1}, we note that it is more grounded in the context of ethical hacking through a client-freelancer relationship, whereas the finding in Section \ref{R1.1} relates more to ethical dilemmas faced by employees working in larger companies.

\subsubsection{Defending Against Cyber Attacks}\label{R2.2}
Our participants (n=2) reported ethical concerns and challenges in relation to defending against cyber attacks. Some of our participants were affected by remote attacks (e.g. DDoS) and were concerned about the use of force to defend against cyber attacks. In some cases, our participants reported the need urgently to take down malicious servers that were conducting attacks on their infrastructure. While our participants had an ethical obligation to protect their organisation's infrastructure, they had significant ethical concerns over the use of force to defend against cyber attacks. Participants were concerned that the use of force may create unacceptable risk to violate the rights of innocent individuals and organisations, since the collateral effects of the use of force are unknown. For instance, participant P04, who works as a security engineer at a large company, explained how they dealt with DDoS against their company's servers:

\begin{quote} ``\textit{We get DDoS attacks all the time, these get automatically blocked with our firewall. If the attack is overwhelming, we'd never retaliate. We go through legal means. Sometimes this means sending abuse reports to the host company. Quite recently, we were dealing with a bunch of IP addresses that were being used to facilitate DDoS amplification attacks. We reported them to the host company and they took them down very quickly.}'' -- P04, Security Engineer \end{quote}

Moreover, Security Developer P10 stated that they avoid the use of force to defend against cyberattacks because cyberattacks often involve `spoofing' strategies which make it easy to misidentify the system responsible for the attack.

\subsection{Systems Security}\label{R3}
Our participants reported a number of ethical concerns, dilemmas and challenges in relation to the Systems Security aspects of cyber security which encompasses Cryptography, Operating Systems \& Virtualisation Security, Distributed Systems Security, Formal Methods for Security and Authentication, Authorisation \& Accountability. Participants experience ethical challenges related to cybersecurity resource allocation (\S \ref{R3.1}), and contracting third party security services (\S \ref{R3.2}). 

\subsubsection{Cybersecurity Resource Allocation} \label{R3.1} Our participants (n=4) reported that there was an ethical concern around the tradeoff between security and other functionalities or priorities. They explained that some cybersecurity solutions (i) may consume considerable individual and organisational resources such as time, labour, money, and expertise; (ii) may negatively impact data storage capacities, bandwidth (upload/download) speeds, energy usage, and the usability and reliability of systems. As such, they experienced challenges in making ethical decisions about how to allocate cyber security resources to protect the security of their customers without creating significant burdens (cost, convenience, and functionality). On one side, not having effective cyber security solutions might cause serious harm and damage to users' security. On the other side, extreme cyber security solutions might be unusable or economically unsustainable. For instance, security manager P13 discussed the challenges of balancing password security policies:

\begin{quote} ``\textit{Balancing this is never easy. We made employees change their passwords every 90 days after one employee had their passwords leaked. The point was to improve our security posture, but that didn't work out. Employees were writing their passwords everywhere.}'' -- P13, Security Manager \end{quote}

\subsubsection{Contracting Third Party Security Services}\label{R3.2}
Our participants (n=9) reported that they had ethical concerns about the use of third party security services. Contracting third party security services (for purposes such as security auditing, penetration testing, security) was effective and cost-efficient for our participants. However, they were concerned about the ethical behaviour of third party services, especially those who operate from countries with challenging political and legal climates. Our participants reported that they exercised due diligence to ensure that contracted third parties are responsible and committed to high standards of ethical conduct. However, some participants reported ethical obstacles that occurred from contracting third parties security services. For instance, Product Security Lead P09 who contracted a remote penetration testing service to evaluate the security of a computer-forensic online service reported a potential inappropriate ethical behaviour. 

\begin{quote} ``\textit{We hired this pentesting firm and signed all the paperwork which clearly said that if they find any vulnerabilities, they need to tell us immediately. We discovered later that they had found but failed to report major vulnerabilities in our website. We contacted them multiple times regarding this, and they never got back to us.}'' -- Product Security Lead, P09 \end{quote}

\subsection{Software and Platform Security}\label{R4}

Our participants reported a number of ethical concerns, dilemmas and challenges in relation to the Software and Platform Security aspects of cyber security which encompasses Software Security, Web \& Mobile Security, Secure Software Lifecycle. Participants experience ethical challenges related to disclosure and patching of vulnerabilities (\S \ref{R4.1}), and prioritising vulnerability patching practices (\S \ref{R4.2}).

\subsubsection{Disclosure and Patching of Vulnerabilities} \label{R4.1} Our participants (n=3) reported ethical concerns and tensions between disclosure and patching of vulnerabilities. Our participants had an ethical duty to be transparent about vulnerabilities found on their system -- so that affected parties can make informed decisions. However, they were wary that disclosing vulnerabilities could make it easier for malicious or bad actors to exploit them. As such, they experienced ethical challenges in balancing between the need for security and the need for transparency. For instance, Systems Security Lead P02 stated that their organisation publicly discloses details of vulnerabilities in their products. However, this is done at the discretion of the organisation and based on a coherent ethical judgement about what is best to do, given the facts, options, products and interests at stake. They explained: 

\begin{quote} ``\textit{There is no one-size-fits-all approach. You have to weigh the risks and benefits involved. A common practice is to disclose them after they have been patched. But, if a vulnerability is critical and can cause severe harm, you might need to disclose it as soon as possible.''} -- Systems Security Lead, P02 \end{quote}

\subsubsection{Prioritising Vulnerability Patching Practices}\label{R4.2}
Our participants (n=6) reported that a common ethical challenge arises when prioritising vulnerability patching practices. Our participants reported that they have an ethical duty to respond to discovered vulnerabilities in a timely manner. Due to the high number of vulnerabilities reported, our participants have to prioritise which vulnerabilities and assets to patch and in what order. Our participants revealed that they often interpret vulnerability risk metrics subjectively and raise a possible dilemma when weighing whether to protect the organisation's interests over those of customers. For instance, Information Security Lead P06 revealed that it is ethically challenging to prioritise between patching vulnerabilities that have been actively exploited in the wild and patching vulnerabilities that are affecting company assets. They said: 

\begin{quote}  ``\textit{It is always challenging to choose which vulnerabilities to patch first because they always lack context. But even if you got more context, you have to make difficult decisions. Do you patch a vulnerability that is being exploited and affecting your users' data first? Or do you patch a vulnerability that can expose your employees' details?}'' -- P06, Information Security Lead \end{quote} 

\subsubsection{Disclosing Security Incidents Without Losing Customer Trust}\label{R4.2}

Our participants (n=3) reported ethical challenges, and dilemmas in disclosing security incidents that have already been addressed and no longer pose a security threat to users and customers. Our participants expressed concerns that disclosing resolved incidents may damage customer trust, and could result in reputational damage to the company or the relevant security team.  As such, participants reported carefully considering the ethical implications of disclosing already-solved incidents, and made informed decisions. Patched vulnerabilities were often disclosed but were kept confidential in some cases.
Privacy Engineer P15 explained: 

\begin{quote}  ``\textit{At that point in time you're like should we or should we not tell the end users? Sometimes there's an incident which is not visible to the end user and you are able to detect and fix it internally on your end. And just deciding as in: Should we tell the users that this happened or should we just remain silent? So you're trying to weigh the cost, you're doing the cost-benefit analysis between telling the end-user the impact of the business that's going to bring. So sometimes you can end up just like: Hey, fix it internally but don't even tell the end user, just keep quiet.''} -- P15, Privacy Engineer \end{quote}

\subsection{Infrastructure Security}\label{R5}Our participants reported a number of ethical concerns, dilemmas and challenges in relation to the Infrastructure Security aspects of cyber security which encompasses Applied Cryptography, Network Security, Hardware Security, Cyber Physical Systems and Physical Layer and Telecommunications Security. Participants experienced ethical challenges related to balancing infrastructure security with privacy (\S \ref{R5.1}), and ensuring accountability and responsibility in AI (\S \ref{R5.2}). 

\subsubsection{Balancing Infrastructure Security with Privacy}\label{R5.1} Our participants (n=2) reported that they faced dilemmas in balancing the security of infrastructure systems with the need to maintain the data protection and privacy of their users. Infrastructure security systems often collect and store data for a variety of purposes, such as protecting users, identifying potential threats, tracking system performance, and improving security measures. However, participants were wary of the privacy and ethical ramifications of using the data of the customers. Security Consultant P14 who works as a security consultant for a company that manages internet infrastructure discussed how they take steps to protect the security of their customers without compromising their privacy. They explained: 

\begin{quote}  \textit{``We could scan customer data but we don't want to do that because it involves essentially snooping on customers traffic. On the flip side of that however, if you move the perspective away from the customer's edge access in their homes, the edge of the network, what we will do is identify and be provided with a list from various agencies and commercial and government groups of known bad URLs and IP addresses, and we can actually drop the traffic in the core network. If you imagine our routing and switching network in London, for example, if we see packets come in on our consumer network, we don't care where they've come from, we don't want to know who sent them into the network, we'll see the IP address that we know is bad or the destination is bad, and we drop the packets there. The consumer is protected because they try to go to a website or service, which is malicious, but we didn't do the content filtering at the edge on their premises.}'' -- Security Consultant, P14 \end{quote}

\subsubsection{Ensuring Accountability and Responsibility in AI} \label{R5.2}
Our participants (n=9) reported a number of security concerns in infrastructure security related to ensuring accountability and responsibility for the use of AI in infrastructure security. They explained that existing infrastructure can often be complex, opaque and has the potential for abuse of power and lack of accountability. Moreover, participants indicated that some infrastructure security systems rely on AI surveillance to identify and track potential threats, which has the potential for mass surveillance and violating user privacy without accountability. Other participants raised ethical concerns about potential for bias in AI systems and the lack of transparency in how AI systems make decisions. Cloud Security Lead P05 explained: 

\begin{quote}  ``\textit{Many companies are now using AI in security technologies and then selling these to governments who then integrate them in critical infrastructure. I don't know if people are aware that AI systems can be attacked. And these attacks can have significant severe effects. I think that is one of the biggest ethical concerns in cybersecurity.''} -- P05, Cloud Security Lead \end{quote}  

We note that unlike other findings, this is a primarily security concern related to the exploitation of bias and discrimination in artificial intelligence.

\section{Discussion} \label{discussion}
\subsection{Ethics of Cybersecurity Decision-Making}

To help frame our discussion, we propose that cybersecurity can be defined as \emph{desirable decisions and actions to manage threats, vulnerabilities, and impacts}. While cybersecurity exists within a legal and regulatory framework, this definition helps us to focus on two fundamental concepts that are central to the ethical challenges that cybersecurity professionals can face: what is \emph{desirable} and how are \emph{decisions} made.

Desirable relates to perception and judgement of what is and what is not wanted but, crucially, cybersecurity always has multiple different stakeholders who may have different perceptions and priorities on matters of security and how to control potential problems. This can clearly lead to conflicts that require resolution in order to act, however it is important to remember that cybersecurity has subjective, objective, and affective dimensions that are specific to each relevant stakeholder. Thus, being able to understand the varied nuances and different perspectives between the objective views on security, the subjective interests of multiple stakeholders, and the emotional implications of cybersecurity actions is necessary. We also note that cybersecurity is a highly technical subject and that cybersecurity actions entail the exercise of power and control over others in various contexts. As a result, different stakeholders have varying levels of security understanding, and also hold cultural, social, and normative views which influence what they perceive to be threats, and what are desirable ways of managing this.

Moreover, in cybersecurity the authority to decide on actions is vested in an individual or small group of individuals who usually (but not always) have the accompanying responsibility of ensuring the success of cybersecurity activities. In order to achieve optimum outcomes, these decision-makers need to resolve complex problems that require a balance between the technical aspects of cybersecurity (threat, vulnerability, impact) and the perceptions and interests of multiple different stakeholders (including subjective and affective dimensions), which is guided by overarching principles. Many of these principles are grounded in ``best practice'' in cybersecurity (e.g. the principle of least privilege, defence in depth, separation of duty, etc.), however these principles also encompass ethical values (e.g. beneficence, justice, non-maleficence, explicability, or autonomy as noted in Section \ref{background}).

Any decision in cybersecurity is therefore driven by the ability of a decision-maker to reach a balance between their own and multiple other stakeholder perspectives on how to manage threats, vulnerabilities and impact, which covers areas such as compliance with relevant laws and regulation, being sensitive to contextual issues (such as cultural, social or religious norms), and adhering to fundamental security, privacy and ethical principles. While practitioners can rely on precedent and ``best practice'' to navigate these decisions, it is readily apparent that there are a number of issues that are not covered by this. Specifically, while precedent can provide helpful framing and information pertaining to past decisions, cybersecurity is an evolving problem: new threats and vulnerabilities get discovered, new security technologies and practices are devised, the likelihood of attacks happening can change over time, and so can contextual factors such as attitudes, practices and expectations. Moreover, while there is a strong desire for ``best practice'' or ``tried-and-tested'' security, this typically results in a ``solutioneering'' approach: where a pre-existing cybersecurity solution is applied to multiple different problems and contexts without the understanding required for making appropriate decisions. Furthermore, the insistence on ``tried-and-tested'' security means that innovation is discouraged and improvements are slow to be recognised and adopted.

\subsection{Decisions in Cybersecurity Professions}

As our findings indicate, decisions in cybersecurity can be made by many different individuals across the profession. Using CyBOK, our findings help map out ethical issues that arise in different areas of the cybersecurity profession, and provides a useful starting point for a more systematic exploration. The issues identified in section \ref{findings} highlight several issues that arise when making specific decisions: difficulties in balancing different perspectives on prioritising vulnerability patching, navigating the emotional aspects of customer trust that can be impacted by disclosing information, or resolving conflicting views about the contrasting interests of the business against the interests of the customers.  More research is needed to systematically map the different types of decisions that arise in the pursuit of professional duties, and help to provide a more specific breakdown of the kinds of ethical dilemmas that may arise in each relevant area of the profession. This would be beneficial in cataloguing the variety of security decisions that pose ethical dilemmas, informing curriculum design and educational material for cybersecurity professionals, and providing deeper opportunities to examine cybersecurity ethical issues in more detail. While the ethics of cryptographic research \cite{rogaway2015moral} or vulnerability disclosure \cite{matwyshyn2010ethics} have been highlighted and investigated, we identified a number of other ethical challenges that need greater scrutiny, such as considering the ethical implications of associating with irresponsible third parties (either clients or service providers), possible harm arising from how security information is interpreted by customers, or navigating conflicts between company and customer interests.

One dimension that was not identified explicitly by our participants is the inherent potential for them to be placed in a personal conflict of interest when making such decisions: for example when a possible decision outcome would result in a negative personal impact but a much larger benefit for others, or when weighing the convenience of simply reusing a pre-existing solution compared to a more detailed evaluation of how suitable it is to the context of use.

\subsection{Ethical Cybersecurity AI}
Finally, it is important to recognise the growing concern over the use and applicability of AI technology in cybersecurity. It is foreseeable that this technology will be used in ways to improve the quality, timeliness and relevance of information available to decision-makers and also to help them identify appropriate solutions. It is also likely that AI models will be used and increasingly be relied upon to make decisions about detailed technical or time-critical cybersecurity problems, however a fundamental concern about AI models pertains to their lack of transparency and explicability. AI decision tools will need to account for the complexity of different perspectives, values, principles outlined above, however it is likely that they too will be subject to bias, unfair, and even malicious influence. Moreover, given the complexity and lack of transparency inherent in these approaches, such biases, malicious influences, or conflicts of interest may become harder to identify, challenge, and remediate. 

\subsection{Implications for Professional Practice}
As noted in Section \ref{codesofpractice}, the current codes of practice from professional bodies do not provide very detailed or nuanced guidance to help practitioners navigate ethical dilemmas in the pursuit of their duties. As discussed above, we believe more research is needed to map and explore the ethical challenges that arise in different areas of the cybersecurity profession, however it is challenging to make recommendations for how these problems should be addressed on a more pragmatic level. Addressing the need for greater consideration of ethical issues through legislation is unlikely to succeed due to the complex and varied nature of moral dilemmas which can arise in cybersecurity. However, insights can be drawn from the long history of medical ethics and practice: from the early origins of the Hippocratic Oath to more modern Codes of Medical Ethics. In particular, we note that due to the manner in which the medical profession is regulated (e.g. through the General Medical Council in the UK) violations of such codes of conduct can lead to disciplinary proceedings and result in the loss of a license to practice medicine. As cybersecurity codes of ethics continue to be developed and updated, we think that it is also important to consider and investigate how ethical breaches could lead to disciplinary proceedings and affect professionals' license to practice cybersecurity.

Furthermore, we believe it is crucial for ethical awareness and training to be integrated into and throughout the education of cybersecurity professionals. This should complement work towards embedding ethical and responsible values into the principles, practices and frameworks that collectively make up the cybersecurity profession. 
Both of these recommendations are highly consistent with the aims of CyBOK, and we argue that (1) more work needs to be done to identify, map, and inform the ethical implications of cybersecurity practice in each of the CyBOK Knowledge Areas; (2) greater efforts need to be made to influence the curriculum, education, and training of cybersecurity professionals to cover and help them manage to navigate ethical dilemmas; (3) more should be done to share and learn from how ethical dilemmas have been tackled by practitioners, including making room for more ``honest'' and ``safe'' spaces for exchange among security officers in relation to the ethical dilemmas they encounter. Finally, (4) it is imperative for more research to be undertaken into the ethics of cybersecurity AI applications, in order to identify and frame the possible benefits, disadvantages, and dangers of this technology.

\section{Limitations}
Our study has some limitations. 
First, when we engaged with our participants, we devised an interview guide that was informed by two different existing pieces of work: CyBOK and the principlist framework proposed by Formosa et al. \cite{formosa2021principlist}. Despite our best intentions in explaining and grounding our engagement in ethical principles, it is evident that much of this detail was overlooked in favour of discussing example cases of ethical challenges. Upon further reflection, it is clear that a principlist approach to empirical ethical research presents challenges in how to engage with participants, however we believe that such principles are very helpful in interpreting the examples and comments given by our participants. In contrast, the detailed breakdown of different areas of cybersecurity from CyBOK provided a much more helpful framing for our participants and proved a useful means of prompting for different experiences of ethical issues across the whole spectrum of cybersecurity practice.

Second, common to all qualitative studies, researcher bias is a concern. Both researchers were trained to conduct research interviews, taking care to avoid leading questions, and ensuring that participants felt comfortable to respond to questions. The researchers avoided interrupting participants, and probed for more information when required. 

Third, given that cybersecurity and ethics are sensitive topics, participants may have been concerned about sharing information, or otherwise been biased in their responses. In order to allay these concerns, an information sheet was given to all participants which described the details of the research and how their data would be used. Each participant was verbally invited to ask questions about this sheet prior to the interviews starting, and we emphasised that their comments would be anonymised and the interview data encrypted and protected. While it is difficult to avoid biased answers when undertaking interview research, we tried to account for this in our analysis of the data by carefully reviewing the codes and themes we identified. Given that our research was positioned as a high level exploration of ethical issues in different areas of the cybersecurity profession, we did not feel it necessary to take additional steps to avoid biases at this stage.

Fourth, the number of interviews is relatively small, and there may be concerns about the validity or generalisability of these findings more widely. While qualitative studies do not aim for statistical representativeness, they do aim to identify conceptual tendencies that can be considered more widely. In keeping with other forms of qualitative research, our aim here is to gain an in-depth understanding of the participants' views on ethics according to different areas of the cybersecurity profession. It is from this detailed understanding that our themes have emerged, and we argue that these can form the basis of future research aiming to generalise this to wider populations and the broader field of cybersecurity ethics.

Finally, in common with other work in empirical ethics, our qualitative findings are limited to the experiences and views of our participants. In particular, given that many of our participants had years of experience working in industry, it is not likely that they would have benefited from more recent efforts at teaching ethics as part of the curriculum of cybersecurity education. Furthermore, our discussion necessarily incorporates our normative views -- while we have endeavoured to ensure that relevant information from our data and appropriate prior work and relevant theory has been presented, the choice and conclusions we draw are strongly influenced by our views as researchers. 

\section{Conclusion}
In this paper, we have presented a review of existing work in cybersecurity ethics, and the results of our investigation into ``What ethical challenges, insights, and solutions arise in different areas of cybersecurity?' We interviewed 15 cybersecurity professionals to investigate this, and found a number of different challenges which we analysed and mapped to the different areas of the Cybersecurity Body of Knowledge. Our findings highlight the inherent complexity and nuance associated with making decisions in cybersecurity, and we discussed how different objective, subjective and affective perspectives need to be taken into account, together with balancing technical, security, and ethical considerations. We further outlined the challenges for professionals who make such decisions and concluded that future research should investigate how ethical values can be embedded throughout the cybersecurity profession, including its tools, practices, and processes. In addition, it is imperative for ethical training and education to become part of the cybersecurity curriculum in order to provide professional practitioners with the means of identifying and addressing the ethical challenges that will arise in the performance of their duties, and we believe that CyBOK offers a very helpful framework for doing so.

Our participants also reported widespread concerns associated with the use of AI in cybersecurity, and we discussed how the technology offers opportunities to support ethical decision-making by providing greater insights and information into the problems being addressed. We also identified that the challenges of making complex cybersecurity decisions are likely to drive the application and adoption of AI-based decision-making. Given the complexity, subjectivity, and contextual nature of cybersecurity decision-making, and coupled with concerns around lack of transparency, possible biases, and potential for being maliciously influenced, we urge the research community to continue and expand the work undertaken in exploring the ethics of cybersecurity AI.

\bibliographystyle{elsarticle-harv} 

\bibliography{sample-base}





\clearpage
\appendix
\section{Ethics of the Cyber Security Profession: Interview Guide}

\textit{Scoping Questions}
\begin{enumerate}
\item Can you tell us about yourself? How long have you worked in cyber security?
\item What is your role in the company that you work at? 
\item When did you join the company?
\item What are your responsibilities?
\item In which of the following areas of cybersecurity have you had any professional involvement:
	\begin{itemize}
	\item Human, Organisational \& Regulatory Aspects (e.g. risk management, compliance, law, human factors, privacy)
	\item Attacks \& Defences (e.g. malware, forensics, incident management)
	\item Systems Security (e.g. cryptography, virtualisation security, Identity/Authentication/Authorization)
	\item Software and Platform Security (Software Security, Web \& Mobile Security, Software Development Lifecycle)
	\item Infrastructure Security (e.g. Network Security, Hardware Security, Cyber Physical Security, Telecommunications Security) 
	\end{itemize}
\end{enumerate}
\textit{General Ethical Questions}
\begin{enumerate}
\item In your opinion, what are the most important ethical considerations for cyber security professionals?
\item Do you think that cyber security professionals should be held to the same ethical standards as other professions?
\item What challenges or obstacles do cyber security professionals face when trying to adhere to ethical principles?
\item How can we ensure that the cyber security profession is accountable and transparent in its decisions and actions?
\item Do you think that cyber security professionals should be held liable for any harm caused by their activities?
\end{enumerate}
\textit{Professional Ethical Practices }
\begin{enumerate}
\item How do you ensure that your cyber security activities are conducted in an ethical manner? How do you evaluate the ethical implications of a cyber security decision or action?
\item What ethical issues do you encounter in your day-to-day work as a cyber security professional? Have you ever experienced ethical dilemmas in your profession? Can you relate them to the following areas of cybersecurity:
\begin{itemize}
	\item Human, Organisational \& Regulatory Aspects (e.g. risk management, compliance, law, human factors, privacy)
	\item Attacks \& Defences (e.g. malware, forensics, incident management)
	\item Systems Security (e.g. cryptography, virtualisation security, Identity/Authentication/Authorisation)
	\item Software and Platform Security (Software Security, Web \& Mobile Security, Software Development Lifecycle)
	\item Infrastructure Security (e.g. Network Security, Hardware Security, Cyber Physical Security, Telecommunications Security) 
	\end{itemize}
\item Have you ever seen or experienced situations where harm occurs to users, employees, or organisations arising from cybersecurity activities or technologies? How would you handle it? \textbf{(Non-maleficence)}
\item How do you ensure that cyber security activities are conducted with respect for the rights of users and organisations? How do you make sure that your activities are beneficial to humans, promote human well-being, and make our lives better overall? \textbf{(Beneficence)}
\item How do you ensure that your cyber security activities respect human autonomy? How do you make sure that users use informed decisions for themselves about how that technology is used in their lives? \textbf{(Autonomy)}
\item How do you ensure that your cyber security activities promote fairness, equality, and impartiality? How do you make sure that they do not unfairly discriminate, undermine solidarity, or prevent equal access? \textbf{(Justice)}
\item How do you ensure that your cyber security activities are used in ways that are intelligible, transparent, and comprehensible? How do you make sure it is accountable and responsible for its use for cyber security technologies? \textbf{(Explicability)}
\item What do you think the role and implications of AI will be for cybersecurity ethics? 
\end{enumerate}

\end{document}